\documentclass[%
 amsmath,amssymb,
letterpaper,aps,superscriptaddress,floatfix,prb,twocolumn
]{revtex4-1}

\usepackage[colorlinks=true, allcolors=blue]{hyperref}

\usepackage{graphicx}% Include figure files
\usepackage{dcolumn}% Align table columns on decimal point
\usepackage{bm}% bold math
\begin{document}

\preprint{APS/123-QED}

\title{Low-temperature thermal conductivity of Co$_{1-x}$M$_x$Si (M=Fe, Ni) alloys}

\author{Y.V. Ivanov}
%\email{Author@institution.edu}
\affiliation{Ioffe Institute, Saint Petersburg 194021, Russia}
\author{A.A. Levin}
\affiliation{Ioffe Institute, Saint Petersburg 194021, Russia}
\author{S.V. Novikov}
\affiliation{Ioffe Institute, Saint Petersburg 194021, Russia}
\author{D.A. Pshenay-Severin}
\affiliation{Ioffe Institute, Saint Petersburg 194021, Russia}
\author{M.P. Volkov}
\affiliation{Ioffe Institute, Saint Petersburg 194021, Russia}
\author{A.Yu. Zyuzin}
\affiliation{Ioffe Institute, Saint Petersburg 194021, Russia}
\author{A.T. Burkov}
\affiliation{Ioffe Institute, Saint Petersburg 194021, Russia}
\author{T. Nakama}
\affiliation{Department of Science, University of the Ryukyus, Okinawa, Japan}
\author{L.U. Schnatmann}
\affiliation{Institute for Metallic Materials, Leibniz Institute for Solid State and Materials Research, 01069 Dresden, Germany}
\author{H. Reith}
\affiliation{Institute for Metallic Materials, Leibniz Institute for Solid State and Materials Research, 01069 Dresden, Germany}
\author{K. Nielsch}
\affiliation{Institute for Metallic Materials, Leibniz Institute for Solid State and Materials Research, 01069 Dresden, Germany}

\date{\today}

\begin{abstract}
We study the low-temperature electrical and thermal conductivity of CoSi and Co$_{1-x}$M$_x$Si alloys (M = Fe, Ni; $x \leq$ 0.06). 
Measurements show that the low-temperature electrical conductivity of Co$_{1-x}$Fe$_{x}$Si  alloys  decreases at $x > $ 0.01 by an order of magnitude compared with that of pure CoSi. 
It was expected that both the lattice and electronic contributions to thermal conductivity would decrease in the alloys. 
However, our experimental results revealed that at temperatures below 20~K  the thermal conductivity of Fe- and Ni-containing alloys is several times larger than that of pure CoSi. We discuss possible mechanisms of the thermal conductivity enhancement. 
The most probable one is related to the dominant scattering of phonons by charge carriers. 
We propose a simple theoretical model that takes into account the complex semimetallic electronic structure of CoSi with nonequivalent valleys, and show that it explains well the increase of the lattice thermal conductivity with increasing disorder and the linear temperature dependence of the thermal conductivity in the  Co$_{1-x}$Fe$_x$Si alloys below 20~K. 
\end{abstract}

\maketitle

\section{Introduction}
\label{sec:1}

Cobalt monosilicide attracted an increased interest recently due to its unusual electronic topology. 
Its electronic structure contains multifold band crossings with large Chern numbers $\pm $4 and long Fermi arcs connecting the projections of the $\Gamma $ and $R$ points on the surface Brillouin zone \cite{bradlyn2016,chang2017a,tang2017a,pshenay2018a}.
In addition, numerous bands with a nonparabolic dispersion near the Fermi level and the semimetallic type of conductivity lead to appearance of some unusual features in transport properties of this compound and its alloys with other transition metal silicides.
For example, strong quantum oscillations of thermoelectric power and magnetoresistivity with beating pattern, arising due to coexistence of two close Fermi surfaces around the $R$ point, were observed in CoSi \cite{xu2019,wu2019}. 
In the Co$_{0.96}$Fe$_{0.04}$Si alloy, a sharp decrease of resistivity with decreasing temperature below 50~K was found \cite{burkov2017b}. 
This feature can arise because of the weak antilocalization of charge carriers. 
The quantized circular photogalvanic effect was predicted and the photocurrent induced by circularly polarized light was recently measured in CoSi and in isostructural $\beta $-RhSi~\cite{chang2017a,rees2020,ni2020}.

In this article, we present experimental results on thermal and electrical conductivity of CoSi and of Co$_{1-x}$M$_{x}$Si (M = Fe or Ni, $x \leq $ 0.06) alloys at temperatures from 2~K to 300~K and discuss their unusual dependence on temperature and composition,  closely related to the band structure of CoSi. 
Measurements revealed that the low-temperature electrical conductivity of  Co$_{1-x}$Fe$_{x}$Si alloys is by an order of magnitude lower compared with that of CoSi. 
A similar behavior of thermal conductivity was anticipated. 
The alloy scattering of phonons is expected  to suppress the lattice contribution to thermal conductivity. 
The electronic contribution to thermal conductivity of an alloy should be reduced in accordance with Wiedemann-Franz law. 
However, the measured thermal conductivity of the alloys at temperatures below 20 K  is several times larger than that of cobalt monosilicide. 
We analize possible mechanisms which can lead to the observed thermal conductivity and show in this work that the most likely mechanism for increasing thermal conductivity in the alloys is the suppression of dominant phonon-electron coupling in conductors with a very small mean free path of charge carriers \cite{pippard1955,pippard1957}. 
This mechanism not only accounts for the increase of the lattice thermal conductivity of Co$_{1-x}$M$_{x}$Si compared to CoSi, but also explains its nearly  linear temperature dependence in the Co$_{1-x}$Fe$_{x}$Si alloys at low temperatures.

\section{Experimental procedures}
\label{sec:2}

Samples of Co$_{1-x}$M$_x$Si were  prepared by direct melting of stochiometric amounts of components in a furnace  with resistive heating followed by vacuum casting.
The ingots of cylindrical shape were re-crystallized by Bridgeman method with inductive heating.
The phase composition and structure of the samples were controlled by X-ray diffraction (XRD).
For the XRD measurements, plates with size of 10$\times$12 mm$^{\rm 2}$ and a thickness of 2~mm were cut out of ingots and polished. 
Also, prepared ingot samples of each composition were milled to powder with particles of about micron size. 
The XRD patterns were registered by means of the powder X-ray diffractometers Rigaku MiniFlex (Rigaku Corporation, Japan) and D2 (Bruker AXS, Germany) designed in Bragg-Brentano geometry. 

Electrical resistivity and thermal conductivity of the alloys were measured, using  Quantum Design PPMS system at 2 to 300~K.

\section{Results and discussion}
\label{sec:3}

\begin{figure}
 \begin{center}
 \includegraphics[scale=1.18]{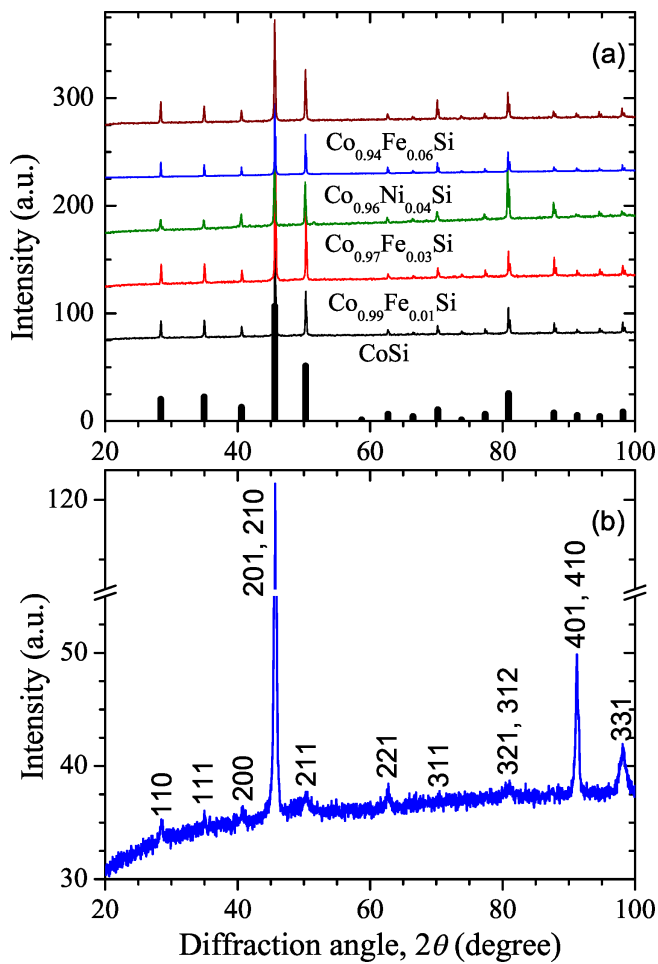}
\end{center}

\caption{(a): XRD patterns of CoSi and Co$_{1-x}$M$_x$Si powder samples and (b): XRD pattern of Co$_{0.96}$Ni$_{0.04}$Si collected from polished surface of the plate sample  ($\Theta $-2$\Theta $ scans, Cu-K$\alpha $ radiation).  For better visualization, the XRD patterns are offset vertically. The bottom histogram in (a) illustrates the theoretical 2$\Theta $ angle positions and intensities of CoSi reflections according to PDF-2 card. Miller indices {\it hkl} of selected observed reflections are indicated in (b).}
\label{xray}
\end{figure}

The powder XRD patterns, shown in Fig.~\ref{xray}(a) for all alloy powders, correspond to CoSi crystal structure (space group  P2$_{\rm 1}$3 (198), Powder Diffraction File-2 (PDF-2) card 01-079-8014) \cite{pauling1948,vandermarel1998,demchenko2008,pdf}.
No foreign phases were detected by the X-ray analysis.  
Diffraction patterns, collected from polished bulk sample surface (see Fig.~\ref{xray}(b) as an example)  are characterized by reflections in the same positions as in powder samples, but with a significant difference in the intensity of reflections, indicating that the ingot samples are strongly textured polycrystals.

\begin{figure}
\centering
\begin{center}
 \includegraphics[scale=1]{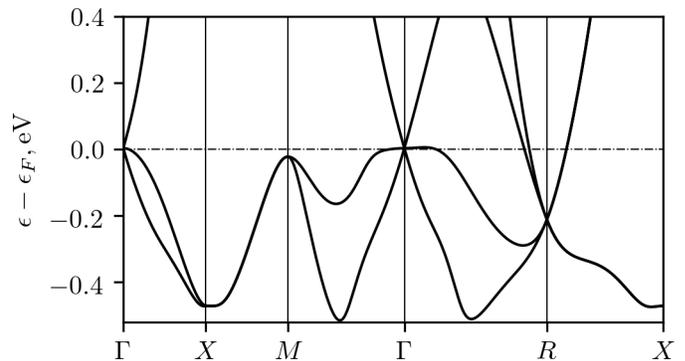}
\end{center}
\caption{\label{fig1}The band structure of CoSi near the Fermi level calculated using the gradient-corrected density functional (GGA-PBE) approximation.}
\end{figure}
The band structure of cobalt monosilicide was calculated using the gradient-corrected density functional (GGA-PBE) approximation \cite{chang2017a,tang2017a,pshenay2018a,pshenay2018b}. 
The structure in a vicinity of Fermi energy, calculated without spin-orbit coupling (SOC) is shown in Fig.~\ref{fig1} \cite{pshenay2018b}. 
The band structure features several extrema and topologically non-trivial band touching nodes near the Fermi level. 
The nodes are located at two time-reversal invariant points in the Brillouin zone ($\Gamma $ and $R$ points) and carry nonzero topological charges. 
Without SOC, the topological charges at these points have magnitude 2 and opposite signs~\cite{tang2017a}.
With the account of spin-orbit splitting~\cite{tang2017a,pshenay2018a}, the multiplet located at the $\Gamma $ point is four-fold degenerate, while the node at the $R$ point is six-fold degenerate, and 
they carry total topological charges of magnitude 4 and opposite signs.
Due to small magnitude of spin-orbit splitting, it was not taken into account in the present work.

The node at the $\Gamma $ point is located very close to Fermi energy, whereas the multiplet at $R$ point is situated at about 200~meV below (see Fig.~\ref{fig1}). 
Near the centre of the cubic Brillouin zone ($\Gamma $ point) there are the flat heavy hole band and the Dirac-like bands with linear dispersion.
At the vertices of the Brillouin zone ($R$ points) the nodal point is located too far below the Fermi level and should not have a direct effect on the low-temperature electrical conductivity and electronic thermal conductivity athough the $R$-electrons give main contribution to electronic transport \cite{pshenay2018b}. 
The electronic band structure around the $R$ point in a vicinity of the Fermi energy consists of two nearly coinciding bands.
In our further analysis we will characterize them by their averaged parameters.
In addition, at the $M$ point there are bands, which are located below the Fermi level.
These states are completely filled in CoSi and do not contribute to electron transport at low temperatures (the maximum at the $M$ point shifts lower with inclusion of  many body G$_{\rm 0}$W$_{\rm 0}$ corrections \cite{pshenay2018a}). 
However, the contribution of the $M$ extrema to electron transport in the Co$_{1-x}$Fe$_{x}$Si alloy, whose Fermi level is  lower than that of CoSi, is unclear. 
In what follows, we will not take into account the states near the $M$ points.

\begin{figure}[h]
\centering
\begin{center}
 \includegraphics[scale=1.1]{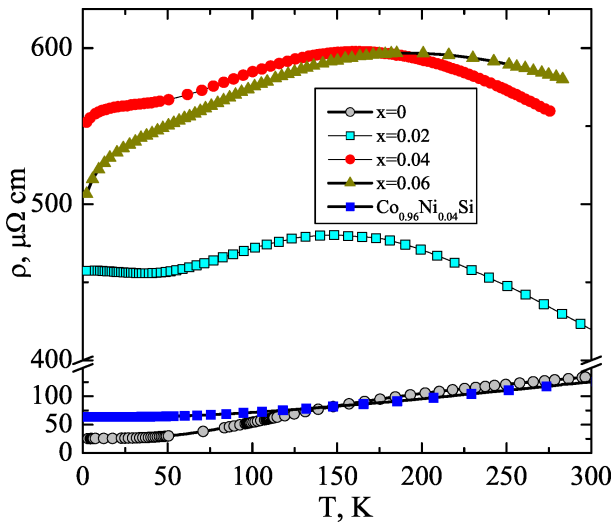}
 \end{center}
\caption{\label{fig2} Temperature dependence of the electrical resistivity of CoSi, Co$_{1-x}$Fe$_x$Si and Co$_{0.96}$Ni$_{0.04}$Si alloys.}
\end{figure}

The temperature dependences of electrical resistivity $\rho$ of CoSi and Co$_{1-x}$M$_{x}$Si are shown in Fig.~\ref{fig2}. 
The temperature dependent resistivity of CoSi has typical metallic character. 
The residual resistivity of CoSi samples is usually quite high, from 10 to 100 ~$\mu \Omega\, $cm \cite{burkov2017b,stishov2012,wu2019}.
This is likely connected with the high sensitivity of the resistivity of CoSi to small deviations from stoichiometry and to intrinsic structural defects.
For our knowledge, the lowest reported  value of the residual resistivity of about 5~$\mu\Omega\, $cm has been measured for single crystal, grown from Te-flux \cite{xu2019}.
Both the resisitivity magnitude and its temperature depencence of Co$_{1-x}$Fe$_x$Si alloys are very strongly dependent of the iron content. 
The residual resistivity increases by about an order of magnitude in alloys with Fe content $x\ge 0.02$. 
The temperature dependences of the alloy resistivity are nonmonotonic. 
The resistivity increases with increasing temperature at low temperatures and then decreases at higher temperatures.
At low temperatures, the temperature dependence of resistivity of the Co$_{1-x}$Fe$_x$Si alloys also undergoes a qualitative transformation with Fe content, showing a Kondo-like minimum for $x\le 0.02$, followed by weak antilocalization-like variation for $x=0.04$ and $x=0.06$.

Some important features of electronic transport  of CoSi and its dilute alloys can be explained on the basis of the ab-initio band structure within energy-dependent relaxation time and rigid band approximations~\cite{pshenay2018b,antonov2019,ovchinnikov2019,xia2019}. 
The Dirac-like band (see Fig.~\ref{fig1}) with linear dispersion at the $\Gamma $ point has relatively small density of states (DOS). 
The relaxation rate of charge carriers in these states is large due to interband scattering into the flat hole band~\cite{pshenay2018b}. 
Therefore, these states make little contribution to electron transport. 
The contribution of heavy holes is also not large because of their small mobility. 
The calculations show that degenerate $R$ electrons make the main contribution to electron transport in CoSi at room temperature and are responsible for the metallic conductivity~\cite{pshenay2018b}. 
According to ab-initio calculations, the Fermi level in Co$_{0.96}$Fe$_{0.04}$Si solid solution is located at about 60~meV below the Fermi energy of CoSi\cite{antonov2019a}. 
This shift of the Fermi level leads to a decrease of $R$-electron concentration. 
In addition, their scattering rate increases due to increasing disorder and intervalley scattering into the flat band with large DOS. 
Therefore, the electrical resistivity of the alloy at low temperatures exceeds by an order of magnitude the resistivity of CoSi (see Fig.~\ref{fig2}) and reaches 550~$\mu \Omega\,$cm at  2~K. 
\begin{figure}[h]
 \begin{center}
 \includegraphics[scale=0.8]{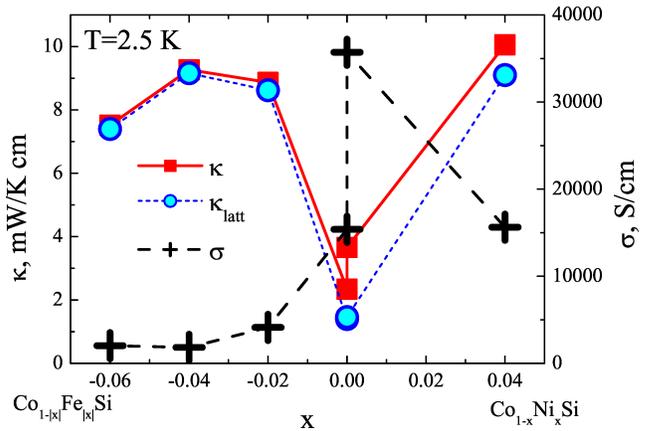}
\end{center}
\caption{Thermal conductivity ($\kappa $), lattice thermal conductivity ($\kappa_{\text{latt}}$) and electrical conductivity ($\sigma $) of CoSi and Co$_{1-x}$M$_x$Si alloys at T=2.5~K vs composition.} \label{r&k-x}
\end{figure}

In alloys with Ni the Fermi level moves to higher energies, increasing the electron concentration.
According to calculations, in the Co$_{0.96}$Ni$_{0.04}$Si alloy, the heavy hole band at the $\Gamma $ point is located at about 200 meV below Fermi energy, therefore the main contribution to electronic transport give the electron pockets around $\Gamma $ and $R$ points \cite{antonov2019a}.
The residual resistivities of the alloy and CoSi are comparable (see Fig.~\ref{fig2}), because the decrease of the relaxation time of $R $-electrons is partly compensated by the increase of their concentration and by the appearence of light $\Gamma $-electrons.

It has been shown that the calculations within the rigid band  and energy-depend relaxation time approximations  reproduce qualitatively the temperature variation of the resistivity and Seebeck coefficient of pure CoSi and diluted Co$_{1-x}$M$_{x}$Si (M=Fe, Ni) alloys at  temperatures above about 50~K~\cite{pshenay2018, antonov2019a}.
This indicates that the band stricture model and the approximations are adequate for interpretation of experimental results on transport properties of CoSi and the alloys.

The thermal conductivity depends on composition and temperature  in an unexpected way.
Figure~\ref{r&k-x} shows the electrical conductivity $\sigma $, the total thermal conductivity $\kappa $ and the lattice thermal conductivity $\kappa _{\rm latt}$ of CoSi and Co$_{1-x}$M$_{x}$Si alloys in dependence on composition at 2.5~K.
The lattice contribution was estimated by Wiedemann-Franz law using the standard value for  Lorenz number $L_0$ = 2.44$\times$10$^{-8}$~W$\Omega$K$^{-2}$.
The first surprising result is that the lowest thermal conductivity has pure CoSi, which in the same time has the highest electrical conductivity. 
The lattice thermal conductivity of both Fe-containing and Ni-containing alloys at low temperatures is by about 5 times higher than $\kappa _{\rm latt}$ of CoSi.
Note, the total thermal conductivity of CoSi and of Co$_{0.96}$Ni$_{0.04}$Si alloy at low temperatures has considerable electronic contribution, whereas the electronic thermal conductivity of Co$_{1-x}$Fe$_{x}$Si alloys is very small.
Another interesting feature relates to CoSi: we have two samples of CoSi, one with residual resistivity $\rho _0$ of 28~$\mu \Omega $cm and second with 65~$\mu \Omega $cm. 
This diffrence comes from small deviations from exact stoichimetry and corresponding structural defects.
The total thermal conductivity of these two samples of CoSi at 2.5~K is also considerably different: 3.66~mWK$^{-1}$cm$^{-1}$ and  2.34~mWK$^{-1}$cm$^{-1}$, respectively, see Fig.~\ref{r&k-x}.
However the $\kappa _{\rm latt}$ of these samples is almost the same.
This means that almost all difference in $\kappa $ of these two samples of CoSi comes from electronic contribution.
On the other hand, $\kappa _{\rm latt}$ of the Co$_{0.96}$Ni$_{0.04}$Si alloy with $\rho _0=64\,\mu \Omega\, $cm (which is very close to $\rho _0$ of less pure CoSi) is 6 times larger than $\kappa _{\rm latt}$  of CoSi (9.1~mWK$^{-1}$cm$^{-1}$ vs 1.46~mWK$^{-1}$cm$^{-1}$).

\begin{figure*}
\centering
\includegraphics[width=0.48\textwidth]{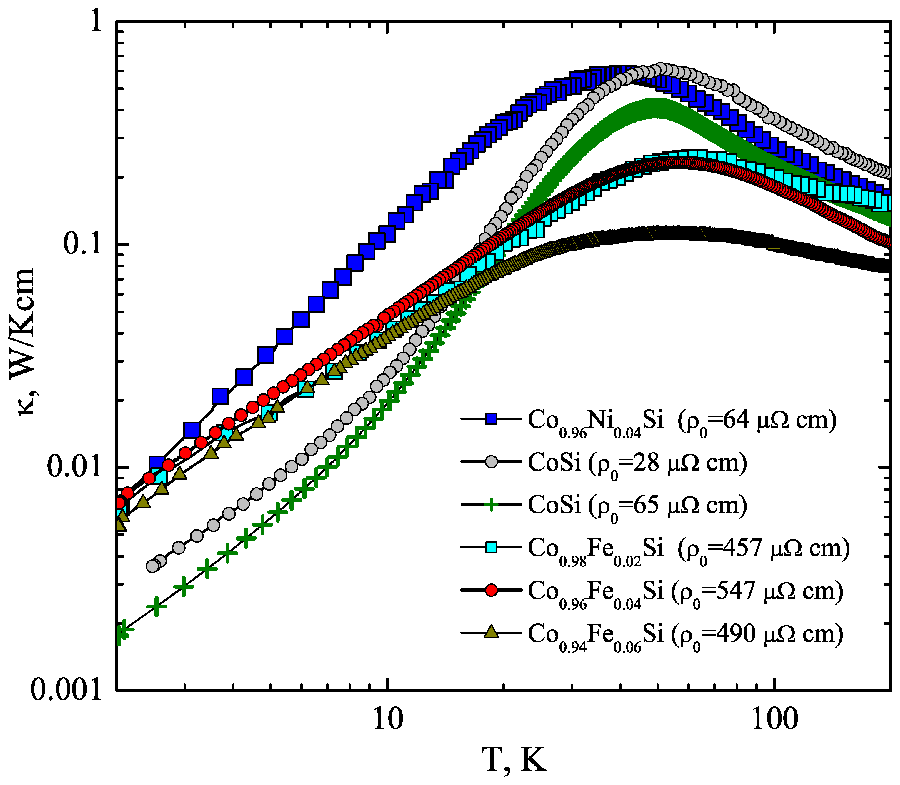}
\includegraphics[width=0.48\textwidth]{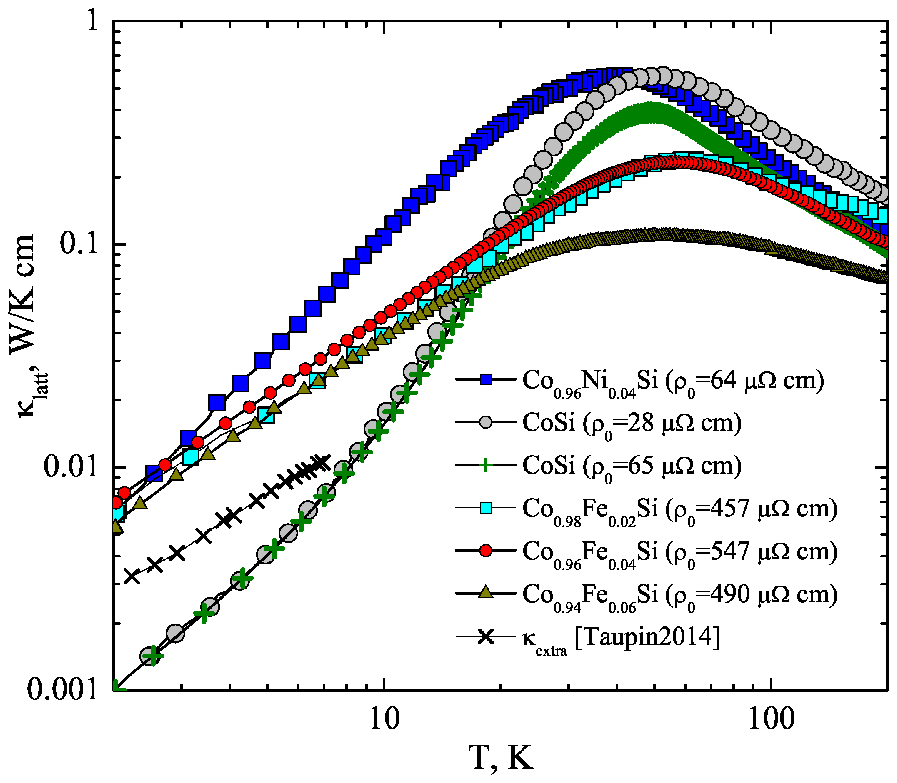}
\caption{\label{fig3}Total (left) and lattice (right) thermal conductivity of CoSi and Co$_{1-x}$M$_x$Si alloys as a function of temperature.  The paramagnon contribution to the thermal conductivity of UCoGe, according to the results of Taupin et al.~\cite{taupin2014}, is also shown.}
\end{figure*}

The temperature dependences of $\kappa$ and $\kappa _{\rm latt}$ of the compounds are shown in Fig.~\ref{fig3}(a) and Fig.~\ref{fig3}(b), respectively. 
The dependences $\kappa _{\rm latt}(T)$ of the Co$_{1-x}$Fe$_{x}$Si alloys are close to linear in the range from 2 to 20~K (for example, $\kappa _{\rm latt}(T)\propto T^{1.19}$ in Co$_{0.96}$Fe$_{0.04}$Si), whereas the thermal conductivity of CoSi and of the Co$_{0.96}$Ni$_{0.04}$Si alloy changes more rapidly with temperature. 
Note that at high temperatures above about 50~K, $\kappa $, like $\sigma $, decreases with the substitution of Co by Fe or Ni. 

The large, in comparison with CoSi, low-temperature thermal conductivity of the Co$_{1-x}$M$_{x}$Si alloys naturally raises the question about an origin of this enhancement. 
Estimates based on Wiedemann-Franz law show that the electronic contribution to the thermal conductivity of the Co$_{1-x}$Fe$_{x}$Si alloys is negligible, while it amounts at low temperatures to about 10\% in the Co$_{0.96}$Ni$_{0.04}$Si alloy and 60\% in the more pure CoSi.
The non-magnetic state of the alloys under study excludes a contribution of magnons to thermal conductivity \cite{sato1955,kumar1982,boona2014}. 
However, in alloys, containing more than 20\% of Fe, the helical magnetic structure is formed at low temperatures \cite{grigoriev2007a}. 
Moreover, comparatively small deviation of stoichiometry can lead to stabilization of magnetic order in CoSi \cite{balasubramanian2020}.
Therefore, one can speculate that the diluted Fe-containing alloys are nearly magnetic semimetals with enhanced itinerant spin fluctuations (paramagnons) which can contribute to their thermal conductivity. 
There are two relevant mechanisms. 
The scattering of electrons by paramagnons can reduce the electronic contribution to thermal conductivity \cite{schindler1967,ueda1975,gratz1995}, which is in any case small in the alloys and would result in oppposite effect, i.e. in a reduction of $\kappa $ in the alloys. 
On the other hand, the itinerant paramagnons can carry heat and directly contribute to thermal conductivity. 
There are few works, where this effect was apparently detected  and more research is needed \cite{lang1977,taupin2014}. 
For example, the paramagnon contribution to the thermal conductivity of UCoGe, is linear in the temperature, and, according to Ref.~\cite{taupin2014}, may have the right magnitude, see Fig.~\ref{fig3}.
Moreover, magnetic measurements show that CoSi, being diamagnetic at high temperatures, has large paramagnetic contribution to the susceptibility at low-temperatures, supporting presence of paramagnetic fluctuations \cite{amamou1972,stishov2012,narozhnyi2013}.
This also is in accord with the changes of the electronic structure of CoSi upon alloying with Fe: in the Fe-containing alloys the Fermi energy decreases entering deeper into the flat valence band at the $\Gamma $ point.
This leads to a rapid increase of DOS and moves electron system closer to Stoner criterion for itinerant magnetism~\cite{pshenay2018,moriya85}.
However this mechanism should not work in case of Ni-containing alloys: in these alloys the Fermi energy increases entering in the highly dispersive region of the electronic structure with comparatively low DOS~\cite{pshenay2018}.
But, as it can be seen in Fig.~\ref{r&k-x}, the lattice thermal conductivity of the Co$_{0.96}$Ni$_{0.04}$Si alloy is also larger than that of CoSi and is comparable to the thermal conductivity of Fe-containing alloys.
There is also an important difference: in contrast to the latter, the thermal conductivity of the Co$_{0.96}$Ni$_{0.04}$Si alloy varies with temperature close to $T^2$.
Therefore, the mechanism of the enhancement of the thermal conductivity  in Co$_{0.96}$Ni$_{0.04}$Si clearly does not fit into paramagnon mechanism.

Considering another possible mechanisms of the thermal conductivity at low temperatures, we can safely exlude the umklapp processes, which can not have a noticeable effect on the phonon transport in CoSi and its alloys due to high Debye temperature (625~K for CoSi~\cite{petrova2010}). 
The phonon scattering by point defects, dislocation strains and grain boundaries also cannot dominate. 
All these processes lead to the $T^{-1}$, $T^2$ and $T^3$ dependences of thermal conductivity~\cite{klemens1958,ziman1960}, respectively, and decrease the thermal conductivity of alloys, when structural disorder increases.

In metals at low temperatures the most important relaxation channel for phonons and acoustic waves is interaction with conduction electrons. 
This relaxation mechanism has been extensively investigated.
Particularly, A.B. Pippard demonstrated~\cite{pippard1955} that attenuation of acoustic waves in metals weakens when the parameter $ql$ decreases well below 1 (here $q$ is the wave number of a lattice vibration, and $l$ is the mean free path of conduction electrons). 
The condition $ql \ll $1 is satisfied at sufficiently low temperatures (small $q$) and for conductors with large residual resistivity (small $l$).
The weakening of the attenuation results in an enhancement of $\kappa _{latt}$.
Additionally, the dependence $\kappa_{\text{latt}}(T)$ in this limit approaches to a linear dependence~\cite{pippard1957}.
The condition $ql \ll $1 can be satisfied in Co$_{1-x}$Fe$_{x}$Si alloys at low temperatures, where the residual resistivity reaches extraordinarily high (for metals) values.
At the same time, the thermal conductivities of these alloys clearly display nearly linear temperature dependences.
Therefore we will discuss this mechanism in more details below.

\section{Evaluation of low-temperature thermal conductivity in semimetals}
\label{sec:4}

The lattice thermal conductivity at low temperatures is described by the expression~\cite{holland1963}:
\begin{equation}
 \kappa=\frac{k_B^4T^3}{6\pi^2\hbar ^3}\sum _{\lambda}s_{\lambda}^{-1}\int_{0}^{\infty}\frac{x^4e^{x}}{\tau_{q\lambda}^{-1}(e^{x}-1)^2}dx,
 \label{kappa}
\end{equation} 
where the summation is performed over the branches of acoustic phonons, $\tau_{q\lambda}^{-1}$  is the relaxation rate of the $\lambda $-th phonon mode with wave vector $q$, $s_{\lambda}$ is the sound velocity, $x=\hbar s_{\lambda}q/k_BT$. 
To estimate the phonon relaxation time $\tau_{q\lambda}$ it is necessary to clarify the mechanism of the phonon-electron interaction.

Attenuation of ultrasonic waves in simple metals due to interaction with conduction electrons (attenuation constant $\alpha_{\lambda}=1/s_{\lambda}\tau_{q\lambda}$), was calculated by Pippard \cite{pippard1955}.
This work initiated extensive investigations of the interaction of lattice vibrations with charge carriers in different conductors (see Refs.~\cite{kittel1955,akhiezer1957,mikoshiba1960,pippard1960,steinberg1958,rayne1970,khan1987,kittel1987,blatt1957,weinreich1959,blount1959,mikoshiba1960a,pomerantz1964,sarges1970,sota1982,sota1983,prunnila2005} and references therein). Although, the Pippard's model describes the sound attenuation or phonon relaxation in simple metals with a spherical Fermi surface, whereas the electronic structure of CoSi and its alloys is more complex, we will first try to estimate the thermal conductivity of Co$_{1-x}$Fe$_x$Si following to the original Pippard's theory.

In metals, the hydrodynamic (diffusive) and quantum regimes of the attenuation of sound waves and phonons are distinguished \cite{pippard1955,khan1987}. 
The crossover between these regimes is determined by the magnitude of the product $ql$. 
In the hydrodynamic regime, the phonon wavelength exceeds the electron mean free path ($ql \ll$ 1). 
A long-wavelength phonon or an ultrasonic wave creates local electronic currents in a metal and loses energy due to Joule heat generation. 
In the quantum regime ($ql \gg $ 1), nonequilibrium phonons lose energy due to  absorption of phonons by electrons.

For numerical estimates, we use material parameter values obtained from ab-initio calculations for CoSi and for the representative alloy Co$_{0.96}$Fe$_{0.04}$Si.
These parameters are listed in Table~\ref{tab1}. 
The total and the interband relaxation rates of charge carriers were estimated in Ref.~\cite{pshenay2018b} for short-range point defect scattering, using Bloch wave functions from DFT calculations. 
The calculation revealed the energy dependence of the electronic relaxation rate and the relative contributions of intraband and interband transitions into the electronic relaxation.
The magnitude of the relaxation rates was scaled to fit  the experimental values of the low-temperature electrical conductivity.
\begin{table}
 \caption{Values of the physical parameters of CoSi ($\rho_0=65 \mu\Omega$\,cm) and Co$_{0.96}$Fe$_{0.04}$Si: $n_{\alpha}$, $N_{\alpha}$ and $k_F^{\alpha}$ are the charge carrier concentration, the DOS and the Fermi wave vector (averaged over directions), respectively; $\tau ^{-1}_{\alpha}$ and $\tau ^{-1}_{\alpha\beta}$ are the total and interband relaxation rates; $s_l$ ($s_t$) is the longitudinal (transverse) sound velocity averaged over directions, $d$ is the density of the material. The indexes $\alpha$ and $\beta$ label the charge carrier pockets ($\Gamma$ and $R$).
\label{tab1}}

\begin{ruledtabular}
\begin{tabular}{lcc}
\textrm{Parameter}&
\textrm{CoSi}&
\textrm{Co$_{0.96}$Fe$_{0.04}$Si}\\
\colrule
$n_{R}$ (10$^{19}$cm$^{-3}$) & 15.5 & 6.1\\
$n_{\Gamma}$ (10$^{19}$cm$^{-3}$) & 30 & 155\\
$N_{R}$ (10$^{21}$eV$^{-1}$cm$^{-3}$) & 1.8 & 1.1\\
$N_{\Gamma}$ (10$^{21}$eV$^{-1}$cm$^{-3}$) & 19.9 & 22.8\\
$k^{R}_F$ (10$^7$ cm$^{-1}$) & 1.65 & 1.21\\
$k^{\Gamma}_F$ (10$^7$ cm$^{-1}$) & 2.15 & 3.76\\
$\tau ^{-1}_R$ (10$^{13}$s$^{-1}$) & 0.7 & 4.0\\
$\tau ^{-1}_{R\Gamma}$ (10$^{13}$s$^{-1}$) & 0.4 & 3.4\\
$\tau ^{-1}_{\Gamma}$ (10$^{13}$s$^{-1}$) & 2.8 & 17\\
$\tau ^{-1}_{\Gamma R}$ (10$^{13}$s$^{-1}$) & 0.05 & 0.35\\
$s_l$ (10$^5$cm/s) & 7.5 & 7.5\\
$s_t$ (10$^5$cm/s) & 4.2 & 4.2\\
$d$ (g/cm$^3$) & 6.6 & 6.6\\
\end{tabular}
\end{ruledtabular}
\end{table}
The data presented in the table are in a resonable agreement with available experimental results. 
For example, according to Ref.~\cite{xu2019} the measured electron concentrations in different CoSi samples  vary from 1.05$\cdot$10$^{20}$  to 3.02$\cdot$10$^{20}$ cm$^{-3}$. 
The calculated total concentration of electrons in two bands near the $R$ point $2n_{R}$ = 3.1$\times$10$^{20}$ cm$^{-3}$. 
The Fermi wave vectors of $R$-electrons estimated from an analysis of the oscillations of magnetoresistance~\cite{wu2019} have lengths 1.42$\times$10$^7$cm$^{-1}$ and 1.31$\times$10$^7$cm$^{-1}$.
The calculations give the value $k_{F}^{R}=1.65\times10^7$cm$^{-1}$ averaged over two bands. 
The quantum scattering time of 3.8$\times$10$^{-13}$s was measured for $R$-electrons at low temperatures~\cite{wu2019}, whereas the theoretical value of the relaxation time $\tau _R$ is equal to 1.4$\times$10$^{-13}$s (note that the low-temperature resistivity of our CoSi sample exceeds that of the sample studied in Ref.~\cite{wu2019} by three times).

Using data of Table~\ref{tab1} we estimate $ql$ values for CoSi and for the representative alloy Co$_{0.96}$Fe$_{0.04}$Si.
For  acoustic phonons with $q \sim k_BT/\hbar s_{\lambda}$, in the free electron model, $ql \sim k_BTk^2_F\tau/(\pi^2\hbar^2s_{\lambda}N)$, where $k_F$ is the length of the Fermi wave vector, $\tau$  is the relaxation time for charge carriers and $N$ is the DOS at the Fermi level. 
Substituting the corresponding values from Table~\ref{tab1}, we obtain for $R$-electrons and transverse phonons $ql \sim $ 0.3 and $ql \sim $ 2  at $T=$2~K in Co$_{0.96}$Fe$_{0.04}$Si and CoSi, respectively. 
For $\Gamma$-holes and longitudinal phonons, the values of $ql$ are smaller. 
Thus, $ql <$ 1 in the alloy, i.e. the relaxation of phonons is close to hydrodynamic regime, whereas, in CoSi, the value $ql \sim$ 2  corresponds to the transition regime.

The Pippard's theory~\cite{pippard1955} describes the sound attenuation in simple metals with a spherical Fermi surface. 
The Co$_{1-x}$Fe$_{x}$Si alloys have a more complex electronic band structure, including several pockets of charge carriers.
Therefore, using the free electron model, we transform the original Pippard's relations to exclude parameters related to specific bands.  
In the limit $ql<1$, the relaxation rates of longitudinal and transverse acoustic phonons in metals~\cite{pippard1955} 
\begin{equation}
\frac{1}{\tau_{q\lambda}}\approx\frac{c_{\lambda}\hbar^2n^{2/3}q^2\sigma}{e^2d},
 \label{pippard}
\end{equation}  
where $n$ is the total concentration of charge carriers, $\sigma$ is the electrical conductivity, $e$ is the electron charge, $d$ is the density of the material. 
The numerical constant $c_{\lambda}$ takes the values of 2.55 and 1.91 for longitudinal and transverse phonons, respectively.
At first glance, the presented relation and Eq.~(\ref{kappa}) make it possible to explain qualitatively the experimental temperature dependences of the thermal conductivity shown in Fig.~\ref{fig3}. 
Indeed, the characteristic value of the phonon wave vector $q\propto T$. 
Therefore, the substitution of $\tau_{q\lambda}^{-1}$ in Eq.~(\ref{kappa}) gives a linear temperature dependence of the phonon thermal conductivity for the Co$_{0.96}$Fe$_{0.04}$Si alloy.
Reducing the Fe content in the alloy slowly decreases $n^{2/3}$ and sharply increases the electrical conductivity. 
Consequently, reduction of the Fe content should lead to an increase of the phonon relaxation rate and a decrease of the thermal conductivity at low temperatures in accordance with the experimental results presented in Fig.~\ref{fig3}.  

Unfortunately, this simple theory strongly overestimates the magnitude of the alloy thermal conductivity. 
For example, using maximum of the theoretical concentration of charge carriers $n=n_{\Gamma}+2n_R = $1.67$\times $ 10$^{21}$~cm$^{-3}$ and the experimental value of electrical conductivity of Co$_{0.96}$Fe$_{0.04}$Si, we obtain $\kappa (2\text{K})\approx0.9$~Wcm$^{-1}$K$^{-1}$, whereas the experimental $\kappa $ for this alloy is of the order of 0.01~Wcm$^{-1}$K$^{-1}$.
Therefore, the original Pippard's theory should be modified, to take into account more complex semimetallic electronic structure of CoSi with comparatively low charge carrier concentration.

In many-valley semiconductors and semimetals, there is  additional mechanism for energy transfer from  phonon or ultrasonic wave to  electron system. 
Local strains produced by the acoustic wave shift the Fermi levels of the valleys relative to each other due to difference their deformation potentials. 
In this case, the interband scattering of electrons by impurities provides relaxation of the system to equilibrium \cite{blatt1957}. 
The calculations performed in Ref.~\cite{pshenay2018b} has shown that, in CoSi, the scattering rate of electrons from states near the $R$ point to those near the $\Gamma $ point is large and is comparable with the rate of intraband scattering. 
Therefore, in CoSi and its alloys with Fe, this mechanism can be very efficient.
Ultrasonic attenuation and phonon relaxation rate in many-valley semiconductors and semimetals were calculated in several works \cite{weinreich1959,blount1959,mikoshiba1960a,pomerantz1964,sarges1970,sota1982,sota1983,prunnila2005}. 
However, only the case of equivalent valleys was investigated in most of these articles. 
The only exception is the Ref.~\cite{blount1959}, where the case of two nonequivalent valleys was considered.
However, in that study, it was assumed that the intervalley relaxation rates of charge carriers of both
bands coincide. 
In CoSi and in Co$_{1-x}$Fe$_x$Si alloys the relaxation rates and dispersion relations of charge carriers near the points $\Gamma $ and $R$ are very different. 
A rigorous calculation of the thermal conductivity of the studied compounds is the topic of a separate article. 
Here we consider a simplest model of phonon relaxation in semimetals with different bands.

In this model, we assume that the interband electron scattering between states of two
bands near the $R$ point and states of the flat band near the $\Gamma $ point is the only
relaxation mechanism for phonons. 
We neglect the intraband diffusion of charge carriers. 
This is possible under the condition $\tau _{\alpha \beta}^{-1} \gg q^2D_{\alpha}$, where $\tau _{\alpha \beta}$ is the interband relaxation time of carriers, $D_{\alpha }$ is the diffusion coefficient of carriers in the band $\alpha $ \cite{weinreich1959,prunnila2005}. 
For $R$-electrons of the Co$_{0.96}$Fe$_{0.04}$Si alloy, the interband relaxation rate $\tau _{R\Gamma}^{-1}$ is comparable with the total relaxation rate $\tau _{R}^{-1}$ (see Table~\ref{tab1}). 
Therefore, the above condition can be approximated by the inequality $(ql_R)^2/3 \ll 1$, where $l_R$ is the mean free path of $R$-electrons. This inequality obviously, holds at low temperatures. 
For $\Gamma $-holes, $D_{\Gamma } \ll D_R$ and the above condition also holds.
For CoSi, the above condition is not satisfied and the intraband diffusion of carriers can be significant. 

In the framework of the described model, the phonon relaxation rate is given by the expression (see Appendix for details):
\begin{equation}
 \frac{1}{\tau_{q\lambda}}=\frac{N_RN_{\Gamma}\Phi _{\lambda}^2}{2d s_{\lambda}^2N}\frac{\omega_{q\lambda}^2\tau}{1+(\omega_{q\lambda }\tau)^2},
 \label{phon_tau}
\end{equation} 
where $d$ is material density,  $N_{R(\Gamma)}$ is DOS at the Fermi level in the band
near the $R$($\Gamma $) point (we neglect small differences in parameters of the two bands
near the $R$ point), $N = 2N_R + N_{\Gamma }$, $\omega _{q\lambda} =  s_{\lambda}q$ is the phonon frequency, and 
\begin{equation}
 \tau=\frac{\tau_{\Gamma R}N_R+\tau_{R\Gamma}N_{\Gamma }}{N}
 \label{el_tau}
\end{equation} 
is the effective interband relaxation time of charge carriers. 
The effective deformation potential constants $\Phi _{\lambda}$ are defined as follows:
\begin{equation}
 \Phi _{\lambda}^2 = \langle \{\left(\Xi _{ij}^{R} - \Xi _{ij}^{\Gamma}\right)\left[ e_i(\mathbf{q},\lambda )\hat{q}_j+e_j(\mathbf{q},\lambda)\hat{q}_i\right]\}^2\rangle .
 \label{def_pot}
\end{equation} 
Here $\Xi _{ij}^{R(\Gamma)}$ is the deformation potential tensor for states near the $R(\Gamma )$ point, $e_i(\mathbf{q},\lambda)$ is the components of the unit polarization vector of the phonon mode $(\mathbf{q},\lambda )$, and $\hat{q}_i =q_i/q$.
The angle brackets denote averaging over directions of phonon wave vectors.

The crystal structure of CoSi belongs to the space group 198. 
The little groups of the $\Gamma $ and $R$ points are isomorphic to the cubic point group T~(23). 
Therefore, the deformation potential tensor $\Xi _{ij}$ at the points $\Gamma $ and $R$ must be proportional to the unit tensor, and, at first glance, only phonons with the longitudinal polarization can interact with electrons. 
However, in our case, there are at least two reasons, why transverse phonons should also interact with charge carriers.
First, the tensor $\Xi _{ij}$ for electronic states near the large Fermi surface of Co$_{1-x}$Fe$_x$Si alloys should differ significantly from the
unit tensor since the little groups of wave vectors of these states are subgroups of the space group.   
Second, in the general case, phonon modes are neither purely transverse nor purely longitudinal.
They have mixed polarization, and can interact with electrons even in the case of the isotropic deformation potential.

In order to estimate the lattice thermal conductivity of CoSi and of Co$_{1-x}$Fe$_x$Si alloys
within the framework of the  model, we substitute the relaxation rates (\ref{phon_tau}) into Eq.~(\ref{kappa}), using the material parameters from Table~\ref{tab1}. 
Since the values of the deformation potential constants for CoSi and Co$_{0.96}$Fe$_{0.04}$Si are unknown, 
we consider them as the fitting parameters. 
For rough estimates, we assume that three constants $\Phi _{\lambda }$ have the same value and do not depend on the concentration of Fe in the alloy. 

\begin{figure}
 \begin{center}
 \includegraphics[scale=0.75]{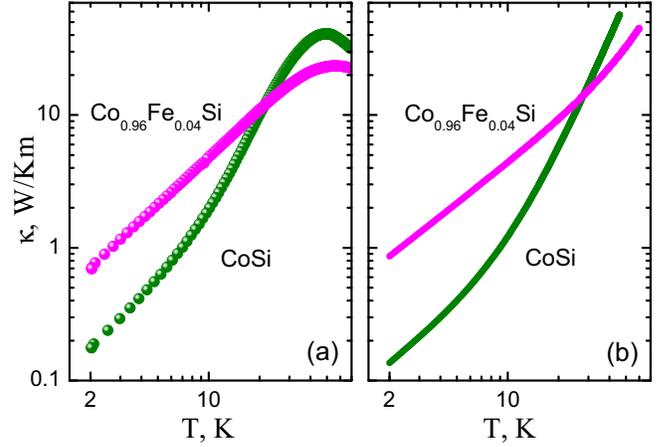}
\end{center}
\caption{Experimental (a) and calculated (b) temperature dependences of the total thermal conductivity of CoSi and the Co$_{0.96}$Fe$_{0.04}$Si alloy.}
\label{Calc}
\end{figure}

The temperature dependencies of the total thermal conductivity of CoSi and the Co$_{0.96}$Fe$_{0.04}$Si alloy calculated by the described method with $\Phi _{\lambda }$ = 2.8~eV are shown in Fig.~\ref{Calc}.
One can see that, in spite of the rough approximations, these dependences qualitatively describe the experimental results at low temperatures. 
The theoretical thermal conductivity increases monotonically with increasing temperature, since we take into account only the phonon-electron coupling and neglect other processes, which dominate at higher temperatures. 

On the experimental part there are at least two apparent inconsistences with the proposed theoretical model.
First is the very close values and the temperature variation of the lattice thermal conductivity of two samples of CoSi with considerably different residual resistivity (65 and 28~$\mu\Omega $cm). 
And the second is the large lattice thermal conductivity of Co$_{0.96}$Ni$_{0.04}$Si alloy with residual resistivity, comparable to that of the less pure CoSi.
Although these experimental facts look like  a real problem to the proposed interpretation, more detailed analysis shows that they  can be understood within the model.

First, the dependence  of $\kappa_{\text{latt}}$ on electronic relaxation time $\tau $ (and consiquently - on residual resistivity), defined by Eqs.~(\ref{kappa}), (\ref{phon_tau}) has a minimum. 
The minimum is located between the values of the effective interband relaxation time of these two CoSi samples, and estimates show that their lattice thermal conductivities should coincide at a temperature of about 5 K. 

In the Co$_{0.96}$Ni$_{0.04}$Si alloy the Fermi level is located above the band crossing at the $\Gamma $ point. 
The residual resistivity of the alloy is relatively small (64~$\mu \Omega $cm) due to high concentration of $R$-electrons. 
Estimates show that due to high concentration of impurities in the alloy its effective interband relaxation time is considerably shorter compared to the relaxation time of CoSi with the residual resistivity of 65~$\mu \Omega $cm. 
In addition, the coefficient $N_RN_{\Gamma}/N$ in Eq.~(\ref{phon_tau}) is small for the alloy. 
Therefore, the phonon relaxation rate (\ref{phon_tau}) in Co$_{0.96}$Ni$_{0.04}$Si at low temperatures should be by almost an order of magnitude smaller than the relaxation rate in CoSi resulting in the enhanced thermal conductivity. 
However, our model cannot explain the relatively large slope of the dependence $\kappa (T)\propto T^{1.8}$, since the decrease of the effective interband relaxation time should lead to the dependence $\kappa (T) \propto T$. 
It can be assumed that a strong decrease of the phonon relaxation rate (\ref{phon_tau}) in the Co$_{0.96}$Ni$_{0.04}$Si alloy leads to a relative increase in the influence of other phonon scattering processes.
For example, an estimate shows that the contribution of the intraband electron diffusion to the phonon relaxation rate is comparable with the contribution of interband electron transitions in the Co$_{0.96}$Ni$_{0.04}$Si alloy. 
Due to the high electronic conductivity and concentration of $R$-electrons, one can try to describe this alloy in the framework of the Pippard model, which gives the dependence~\cite{pippard1957} $\kappa_{\rm latt} (T)=aT + bT^2$, similar to that shown in Fig.~\ref{fig3}. 
Another possible mechanism of phonon dissipation is the phonon scattering by a dislocation strain field. 
Its contribution $ 1/\tau _{q  \lambda } \propto \omega _{q \lambda }$ \cite{klemens1958} together with the considered phonon-electron contribution (\ref{phon_tau}) can give the temperature dependence of the lattice thermal conductivity, similar to the experimental one.
Unfortunately, at present, we cannot unambiguously identify the dominant mechanisms of phonon relaxation in Ni-containing alloys. 
Further research is required.

\section{Conclusions}\label{sec:5}

In conclusion, we have studied the thermal and electrical conductivity of cobalt monosilicide and the alloys Co$_{1-x}$Fe$_x$Si and Co$_{0.96}$Ni$_{0.04}$Si. 
Despite the low electrical conductivity of disordered alloys, their thermal conductivity at low temperature is several times higher than that of CoSi. 
This contradicts to the opinion that the thermal conductivity of a dilute alloy is always lower than that of the pure compound. 
We demonstrated that the enhancement of the low-temperature thermal conductivity in the alloys is related to the weakening of the phonon-electron interaction in compounds with a short electron mean free path compared to the characteristic phonon wavelength.
We estimated the low-temperature thermal conductivity of the studied alloys using a model of phonon-electron coupling, which takes into account the interband scattering of charge carriers between non-equivalent bands. 
This mechanism is effective in CoSi-based semimetals, since the interband and intraband electron-impurity scattering rates are comparable in them. 
The estimates of the thermal conductivity of the investigated semimetals are in a reasonable agreement with the experimental results.  

\section{Acknowledgments}\label{sec:6}
The study was supported by the Russian Foundation for Basic Research, project 18-52-80005 (BRICS).

\appendix
\section{Derivation of Eq.~\ref{phon_tau}}

In this Appendix we present derivation of Eq.~\ref{phon_tau}. 
In semimetals and many-valley semiconductors, the collision term of the Boltzmann equation can be written in the form~\cite{kragler1980}:
\begin{equation}
\begin{split}
&\left(\frac{\partial f_{\alpha}}{\partial t}\right)_{\rm coll}=
\\&-\sum_{\beta}\gamma_{\alpha\beta}\left\lbrace\delta f_{\alpha}({\bf k,r},t)  - \frac{\partial f^0_{\alpha}}{\partial\varepsilon _{{\bf k}\alpha}}\Big[U_{\alpha}({\bf r},t)-\delta\mu_{\alpha\beta}({\bf r},t)\Big]\right\rbrace,
\end{split}
\label{A1}
\end{equation}
where the Greek indices number bands (valleys) of the electron system, $\delta f_{\alpha}({\bf k,r},t)=f_{\alpha}({\bf k,r},t)-f^0_{\alpha}(\bf k)$  is the deviation of the distribution function of charge carriers in the band $\alpha$  from the Fermi-Dirac distribution  
\begin{equation}
 f^0_{\alpha}({\bf k})=\lbrace{exp\left[(\varepsilon _{{\bf k},\alpha}-\mu_0)/k_{B}T\right]+1\rbrace}^{-1},
\end{equation}
$\gamma_{\alpha\beta} = \tau^{-1}_{\alpha\beta}$ is the interband relaxation rate, $U_{\alpha}({\bf r},t)$ is the deformation potential and $\delta\mu_{\alpha\beta}({\bf r},t)= \delta\mu_{\beta\alpha}({\bf r},t)$ is the deviation of the local chemical potential of two corresponding bands from the chemical potential  $\mu_0$ of the system without acoustic waves. 
Integrating the Boltzmann equation with the collision term (\ref{A1}) over $k$-space and omitting the term with electric current (i.e. neglecting the drift and diffusion of charge carriers), we obtain:
\begin{equation}
 \begin{split}
  &\frac{\partial\delta n_{\alpha}({\bf r},t)}{\partial t}=\\
  &-\sum_{\beta \neq \alpha}\gamma _{\alpha\beta}\left\lbrace\delta n_{\alpha}({\bf r},t)+N_{\alpha}\big[U_{\alpha}-\delta\mu_{\alpha\beta}({\bf r},t)\big]\right\rbrace,
 \end{split}
\label{A3}
 \end{equation}
where $\delta n_{\alpha}({\bf r},t)$ is the deviation of the carrier concentration from equilibrium value, and
\begin{equation}
 N_{\alpha} = \int\left(-\frac{\partial f^0_{\alpha}}{\partial \varepsilon _{{\bf k}\alpha}}\right)\frac{2d^3k}{(2\pi)^3}
\end{equation}
is the density of states of the $\alpha $ band  at the Fermi level (with accuracy up to terms proportional to $(k_BT/\mu_0)^2$). 
The term with $\gamma _{\alpha\alpha}$  does not contribute to Eq.~(\ref{A3}) since the intraband scattering does not change the concentration of particles. 
The off-diagonal components of the matrix $\delta\mu_{\alpha\beta}$ are usually determined from the condition that the local particle concentration $n_{\alpha}+n_{\beta}$ and the particle concentration in these two bands in the quasi equilibrium state with the chemical potential $\mu_0+\delta\mu_{\alpha\beta}$ are equal \cite{kragler1980}. 
This condition is applicable only to semiconductors with equivalent valleys. 
In the present case, when $\gamma_{\alpha\beta}\neq \gamma_{\beta\alpha}$, it should be replaced by the condition 
\begin{equation}
 \begin{split}
  &\gamma_{\alpha\beta}\big[\delta n_{\alpha}+N_{\alpha}\left(U_{\alpha}-\delta\mu_{\alpha\beta}\right)\big]=\\
 -&\gamma_{\beta\alpha}\big[\delta n_{\beta}+N_{\beta}\left(U_{\beta}-\delta\mu_{\beta\alpha}\right)\big],
 \end{split}
 \label{A5}
 \end{equation}
i.e., the rate of change of carrier concentration in the band $\alpha$ due to particle scattering between the bands $\alpha$ and $\beta $ should be equal (with the opposite sign) to the rate of change of carrier concentration in the band $\beta$ due to scattering between the same bands.

Below we assume that two bands around $R$ point of the Brillouin zone are equivalent, i.e., they have the same band parameters and the deformation potential does not remove their degeneracy. Moreover, the first principle calculations show that the interband carrier scattering between these two bands is negligible compared to scattering into the flat hole band.
In this case, the Eqs.~(\ref{A3}), (\ref{A5}) can be written in the form: 
\begin{equation}
 \begin{split}
  &\frac{\partial\delta n_R}{\partial t}=-\frac{N_RN_{\Gamma}\Delta U+N_{\Gamma}\delta n_R-N_R\delta n_{\Gamma}}{N\tau}\\
  &\frac{\partial\delta n_{\Gamma}}{\partial t}=\frac{2\left(N_RN_{\Gamma}\Delta U+N_{\Gamma}\delta n_R-N_R\delta n_{\Gamma}\right)}{N\tau},
 \end{split}
\label{A6}
\end{equation}
where $N=2N_R+N_{\Gamma}$  is the total density of states, $\tau $ is the effective interband relaxation time of charge carriers defined by Eq.~(\ref{el_tau}), and $\Delta U=U_R-U_{\Gamma}$.

The long-wavelength acoustic wave causes the displacement of lattice ions:
\begin{equation}
\begin{split}
 &{\bf u}\left({\bf r},t\right)=\\
 &{\bf e}\left({\bf q},\lambda \right)\Big[u\left({\bf q},\lambda\right)e^{i({\bf qr}-\omega t)} + u^*\left({\bf q},\lambda\right)e^{-i({\bf qr}-\omega t)}\Big]
 \end{split}
 \label{A7}
\end{equation}
and the local change of the carrier concentration:
\begin{equation}
 \delta n_{\alpha}({\bf r},t)=\eta _{\alpha}({\bf q},\lambda)e^{i({\bf qr}-\omega t)}+\eta^*_{\alpha}({\bf q},\lambda)e^{-i({\bf qr}-\omega t)},
 \label{A8}
\end{equation}
where ${\bf e}({\bf q},\lambda )$ are the polarization vectors of phonons,  $u({\bf q},\lambda)$ and $\eta _{\alpha}({\bf q},\lambda)$ are the Fourier components of the corresponding variables, $\omega = s_{\lambda }q$. 
Since we consider here the relaxation of a single acoustic wave, the summation over $q$ and $\lambda $ is omitted. 
Substitution of Eqs.~(\ref{A7}) and (\ref{A8}) into the system (\ref{A6}) gives its solution:
\begin{equation}
 \begin{split}
  &\eta _R({\bf q},\lambda)=\frac{N_RN_{\Gamma}}{\left(i\omega \tau -1\right)N}\Big[U_R\left({\bf q},\lambda\right)-U_{\Gamma}\left({\bf q},\lambda\right)\Big]\\
  &\eta _{\Gamma}({\bf q},\lambda)= -2\eta _R({\bf q},\lambda),
 \end{split}
\label{A9}
\end{equation}
where 
\begin{equation}
\begin{split}
 &U_{R(\Gamma )}\left({\bf q},\lambda\right)=\\
 &\frac{iqu({\bf q},\lambda)}{2}\Xi^{R(\Gamma)}_{ij}\Big[e_i(q,\lambda)\hat{q}_j+e_j(q,\lambda)\hat{q}_i\Big],
 \end{split}
 \label{A10}
\end{equation}
$\Xi^{R(\Gamma)}_{ij}$ is the deformation potential tensor, $\hat{q}_i=q_i/q$.

The energy $Q$ transferred from the acoustic wave to carriers per unit volume and per unit time is given by:\cite{sota1982,blount1959}
\begin{equation}
 Q=-2\omega\sum _{\alpha}{\rm Im}\Big[\eta_{\alpha}\left({\bf q},\lambda\right)U^*_{\alpha}\left({\bf q},\lambda\right)\Big].
 \label{A11}
\end{equation}
The phonon relaxation rate is defined as the ratio $Q/E$, where $E=d \langle\left(\partial {\bf u}/\partial t\right)^2\rangle$  is the energy density of acoustic wave, $d$ is the mass density. 
The angle brackets denote time averaging. 
Substitution of Eqs.~(\ref{A7}), (\ref{A9})-(\ref{A11}) into this ratio gives the phonon relaxation rate (\ref{phon_tau}).

\bibliography{}
\end{document}